\documentclass[letters,fleqn,usenatbib]{mnras}

\usepackage{newtxtext,newtxmath}
\usepackage[T1]{fontenc}

\DeclareRobustCommand{\VAN}[3]{#2}
\let\VANthebibliography\thebibliography
\def\thebibliography{\DeclareRobustCommand{\VAN}[3]{##3}\VANthebibliography}

\usepackage{graphicx}	
\usepackage{amsmath}	

\usepackage{fontspec}
\newfontfamily\inuktitutfont[
  Path = ./,
  Extension = .ttf,
  UprightFont = *
]{NotoSansCanadianAboriginal}

\title[Arctic Solar Radio Monitoring with ALBATROS]{Continuous Ultra-Low-Frequency Solar Radio Monitoring with ALBATROS from the High Arctic}

\author[Chokshi et al.]{
Aman Chokshi,$^{1,2}$\thanks{E-mail: aman.chokshi@mcgill.ca}
Mohan Agrawal,$^{1,2}$
Christopher Barbarie,$^{3,1}$
Joëlle-Marie Bégin,$^{4}$
Benjamin Cheung,$^{1,5}$
\newauthor
Hsin Cynthia Chiang,$^{1,2}$
Cherie K. Day,$^{1}$
Eamon Egan,$^{1}$
Stephen Fay,$^{1}$
Nivek Ghazi,$^{6,7}$
Maya Goss,$^{8}$
\newauthor
Lawrence Herman,$^{1}$
Jack Hickish,$^{9}$
Michael Hétu,$^{10}$
Ronniy Joseph,$^{1,2,11}$
Marc-Olivier R. Lalonde,$^{12}$
\newauthor
Tristan Ménard,$^{1}$
John Orlowski-Scherer,$^{1,13}$
Jonathan Sievers,$^{1,2}$
Will Tyndall,$^{1,2}$
Anthony B. Zerafa$^{14}$
\\
$^{1}$Department of Physics, McGill University, Montréal, Québec H3A 2T8, Canada\\
$^{2}$Trottier Space Institute, McGill University, Montréal, Québec H3A 2A7, Canada\\
$^{3}$Department of Electrical, Computer \& Energy Engineering, University of Colorado Boulder, Boulder, CO, United States 80309\\
$^{4}$Department of Physics, Princeton University, Princeton, NJ, United States, 08544\\
$^{5}$Department of Physics \& Astronomy, McMaster University, Hamilton, Ontario L8S 1C7, Canada\\
$^{6}$South African National Space Agency\\
$^{7}$School of Mathematics, Statistics, and Computer Science, University of KwaZulu-Natal\\
$^{8}$Department of Geography, Planning \& Environment, Concordia University, Montréal, QC, Canada H3G 1M8\\
$^{9}$Real-Time Radio Systems Ltd, Bransgore, Christchurch, Dorset, BH23 8AY, UK\\
$^{10}$Kavli Institute for Cosmological Physics, University of Chicago, Chicago, IL, United States, 60637\\
$^{11}$S[\&]T Netherlands, Delft, The Netherlands\\
$^{12}$School of Earth and Space Exploration, Arizona State University, Tempe, AZ, United States, 85287-6004\\
$^{13}$Department of Physics and Astronomy, University of Pennsylvania, Philadelphia, PA, United States, 19146\\
$^{14}$Department of Biology, McGill University, Montréal, Quebec, Canada H3A 1B1\\
}

\date{Accepted XXX. Received YYY; in original form ZZZ}

\pubyear{\the\year{}}

\begin{document}
\label{firstpage}
\pagerange{\pageref{firstpage}--\pageref{lastpage}}
\maketitle

\begin{abstract}
In the Canadian High Arctic, nearly five months of continuous daylight enable uninterrupted low-frequency solar monitoring. We present the first solar science results from the Array of Long Baseline Antennas for Taking Radio Observations from Seventy-Ninth Parallel (ALBATROS). This broadband radio array is designed to explore the largely uncharted radio sky below 30 MHz, where polar ionospheric conditions permit access to frequencies rarely accessible from ground-based sites. Using observations spanning 1–125 MHz, we detect bright solar radio bursts exhibiting complex spectral and polarised structure. The bursts are observed simultaneously by all eight autonomous stations, demonstrating the stability and consistency of the array. Comparison with concurrent soft X-ray measurements reveals a strong temporal correlation between the radio and X-ray emission. These observations establish ALBATROS as a new facility for ultra-low-frequency solar monitoring, opening a new window on solar radio bursts, space weather, and the dynamic heliosphere.
\end{abstract}



\section{Introduction}
\label{sec:intro}

The ultra-low-frequency radio spectrum of the Sun remains one of the least explored windows on solar activity. Solar radio bursts provide direct diagnostics of particle acceleration, shock formation, and magnetic energy release in the corona \citep{WILD_SOLAR_FLARE_1950, NELSON_MELROSE_TYPE_II_1985, REID_RATCLIFFE_TYPE_III_2014, MOROSAN_SOLAR_ACCELERATION_2025}. At frequencies below a few hundred megahertz, the emission is often generated near the local electron plasma frequency and its harmonics, allowing observing frequency to be mapped directly to the ambient electron density and, by extension, to height in the solar atmosphere \citep{VRSNAK_CLIVER_CORONAL_SHOCKS_2008, REID_RATCLIFFE_TYPE_III_2014}. Dynamic spectra therefore trace the propagation of energetic electron beams, coronal shocks, and eruptive structures associated with solar flares and coronal mass ejections. These phenomena drive space weather, which can disrupt satellite operations, radio communications, navigation systems, and power infrastructure on Earth \citep{GOPALSWAMY_SPACE_WEATHER_2006, PULKKINEN_SPACE_WEATHER_2007}.

The spectral regime below $\sim$30 MHz is particularly valuable because it probes the outer corona, where coronal mass ejections, shocks, and energetic particles propagate into the heliosphere and become observable as space-weather drivers \citep{BHONSLE_SOLAR_DECAMETER_1979, CANE_TYPE_III_FLARES_2002, GOPALSWAMY_CME_TYPE_II_2008, WHITE_SOLAR_SPACE_WEATHER_2024}. Yet this frequency range remains difficult to access from the ground, as the terrestrial ionosphere becomes increasingly opaque and variable at low frequencies. Observations near the ionospheric cutoff are therefore possible only intermittently and under favorable ionospheric conditions \citep{FAINBERG_STONE_SATELLITE_TYPE_III_1974, KLEIN_LUNAR_RADIO_2012}, or from space-based observatory concepts and missions such as SunRISE \citep{KASPER_SUNRISE_2022}.

The Canadian High Arctic offers a rare opportunity to access this largely unexplored spectral window. Near the geomagnetic poles, the ionospheric electron density can fall to unusually low values \citep{GRUNEY_THEMENS_IONO_2025}, reducing the local plasma frequency and enabling observations at frequencies that are normally inaccessible from most ground-based sites \citep{BJOLAND_POLAR_IONO_2021}. During summer, the Sun remains continuously above the horizon for nearly five months. While global networks such as e-Callisto \citep{BENZ_ECALISTO_2009} achieve near-continuous coverage by combining observations from geographically distributed observatories, an Arctic observatory can monitor the Sun from a single site for months at a time.

The Array of Long Baseline Antennas for Taking Radio Observations from Seventy-Ninth Parallel (ALBATROS) is designed to exploit this unique observing environment. Located on Umingmat Nunaat\footnote{The Inuktitut name for Axel Heiberg Island: {\inuktitutfont ᐅᒥᖕᒪᑦ ᓄᓈᑦ}} (Axel Heiberg Island, Nunavut) in the Canadian High Arctic, ALBATROS is a broadband radio array that provides continuous dynamic spectroscopy over 1–125 MHz in an exceptionally radio-quiet environment. 

In this Letter, we present the first solar science results from ALBATROS. Section~\ref{sec:instrument} briefly describes the instrument and its Arctic observing geometry. Section~\ref{sec:results} presents representative solar radio bursts, including comparisons with Geostationary Operational Environmental Satellite (GOES) soft X-ray observations, full-Stokes dynamic spectra, and simultaneous detections across all eight stations. We discuss the implications of these observations for ultra-low-frequency solar astronomy and future space-weather studies in Section~\ref{sec:discussion}.

\section{The ALBATROS Array}
\label{sec:instrument}

The ALBATROS array \citep{CHIANG_ALBATORS_2020} comprises eight autonomous radio stations distributed across baselines ranging from 175~m to 8.7~km on Axel Heiberg Island. Each station is equipped with a Long Wavelength Array (LWA) active crossed-dipole antenna \citep{HICKS_LWA_ANTENNA_2012} coupled to a custom front-end and digital back-end. ALBATROS records full-Stokes dynamic spectra spanning 1--125~MHz with 61~kHz frequency resolution and 6.44~s time resolution. Raw baseband voltages are additionally archived at a native time resolution of 16~$\mu$s in either 30~MHz (1-bit) or 8~MHz (4-bit) modes, enabling subsequent offline correlation and interferometric analyses.

\begin{figure}
    \centering
    \includegraphics[width=\columnwidth]{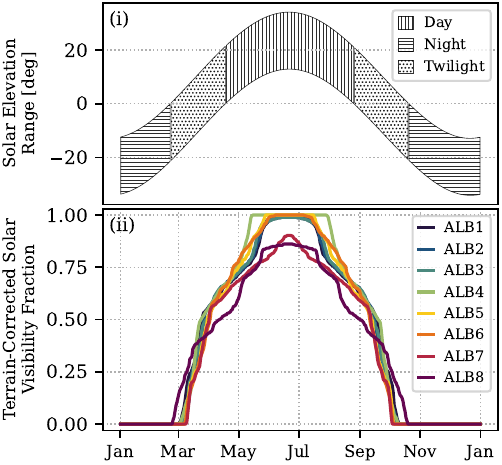}
    \caption{
        Seasonal solar visibility at the ALBATROS site in the Arctic. (\textit{i}) Daily range of solar elevation throughout the year, highlighting periods of continuous daylight, twilight, and polar night. (\textit{ii}) Daily fraction of the Sun visible above the local terrain horizon for each of the eight ALBATROS stations. During summer, nearly five months of continuous daylight enable uninterrupted solar monitoring.
    }
    \label{fig:solar_visibility}
\end{figure}

The array is situated an exceptionally radio-quiet environment \citep{DYSON_MARS_RFI_2021} and within a few degrees of the geomagnetic pole, where the ionospheric plasma frequency can occasionally fall below 10MHz  \citep{BJOLAND_POLAR_IONO_2021, GRUNEY_THEMENS_IONO_2025}. Figure~\ref{fig:solar_visibility}, computed using \textit{solshade} \citep{CHOKSHI_SOLSHADE_2026}, summarizes the annual solar visibility at the ALBATROS site. The upper panel shows the seasonal variation in solar elevation, while the lower panel presents the fraction of the solar disk visible above the local terrain horizon for each station. Although local topography introduces modest variations between stations, the Sun remains visible for nearly the entire summer season.

\section{First Solar Radio Observations}
\label{sec:results}

\subsection{Dynamic Spectra and Soft X-Ray Light Curves}

\begin{figure}
    \centering
    \includegraphics[width=\columnwidth]{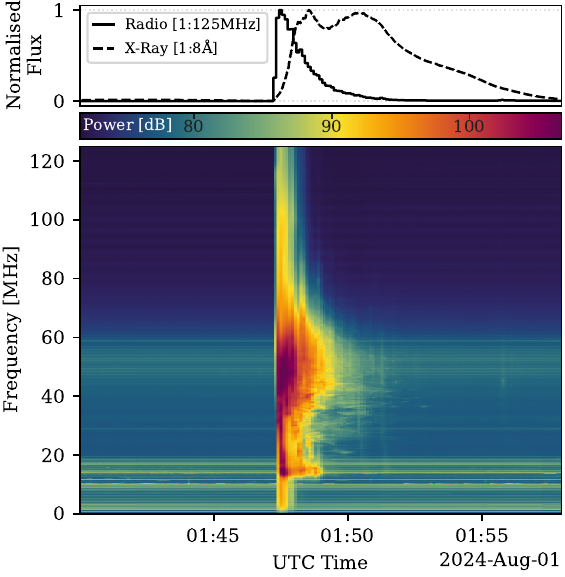}
    \caption{
        Representative solar radio burst observed with ALBATROS.
        The upper panel compares the normalised ALBATROS integrated radio light curve (1--125 MHz) with the simultaneous GOES 1--8 \AA\ soft X-ray flux measurement. The lower panel shows the corresponding dynamic spectrum, revealing broadband emission extending from the ionospheric cutoff to over 100 MHz. The strong temporal association confirms the solar origin of the burst.
    }
    \label{fig:goes_comparison}
\end{figure}

\begin{figure}
    \centering
    \includegraphics[width=\columnwidth]{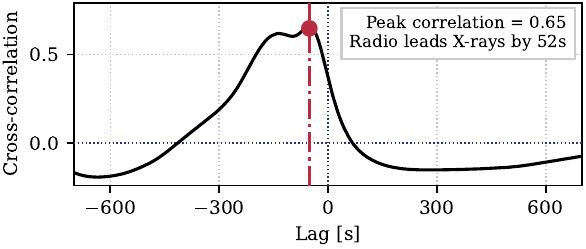}
    \caption{
        Cross-correlation between ALBATROS integrated radio light curve and GOES soft X-ray flux for the event shown in
        Figure~\ref{fig:goes_comparison}. The strong peak demonstrates a close temporal relationship between the radio and X-ray emission.
    }
    \label{fig:radio_xray_correlation}
\end{figure}

Figure~\ref{fig:goes_comparison} shows an illustrative solar radio burst observed with ALBATROS on 2024 August 1, associated with an M6.3-class flare from Active Region 13773. The dynamic spectrum reveals bright broadband emission extending from the local ionospheric cutoff to above 100 MHz, with pronounced fine structure, rapid temporal variability, and a morphology consistent with a Type III solar radio burst. The upper panel compares the frequency-integrated ALBATROS light curve with the 1--8 \AA\ soft X-ray flux measured by the Extreme Ultraviolet and X-ray Irradiance Sensors (EXIS) aboard GOES \citep{MACHOL_GOES_EXIS_2020}. Figure~\ref{fig:radio_xray_correlation} shows the cross-correlation between the radio and X-ray light curves, revealing a strong peak near zero lag.

\subsection{Full-Stokes Polarimetry}

\begin{figure*}
    \centering
    \includegraphics[width=\textwidth]{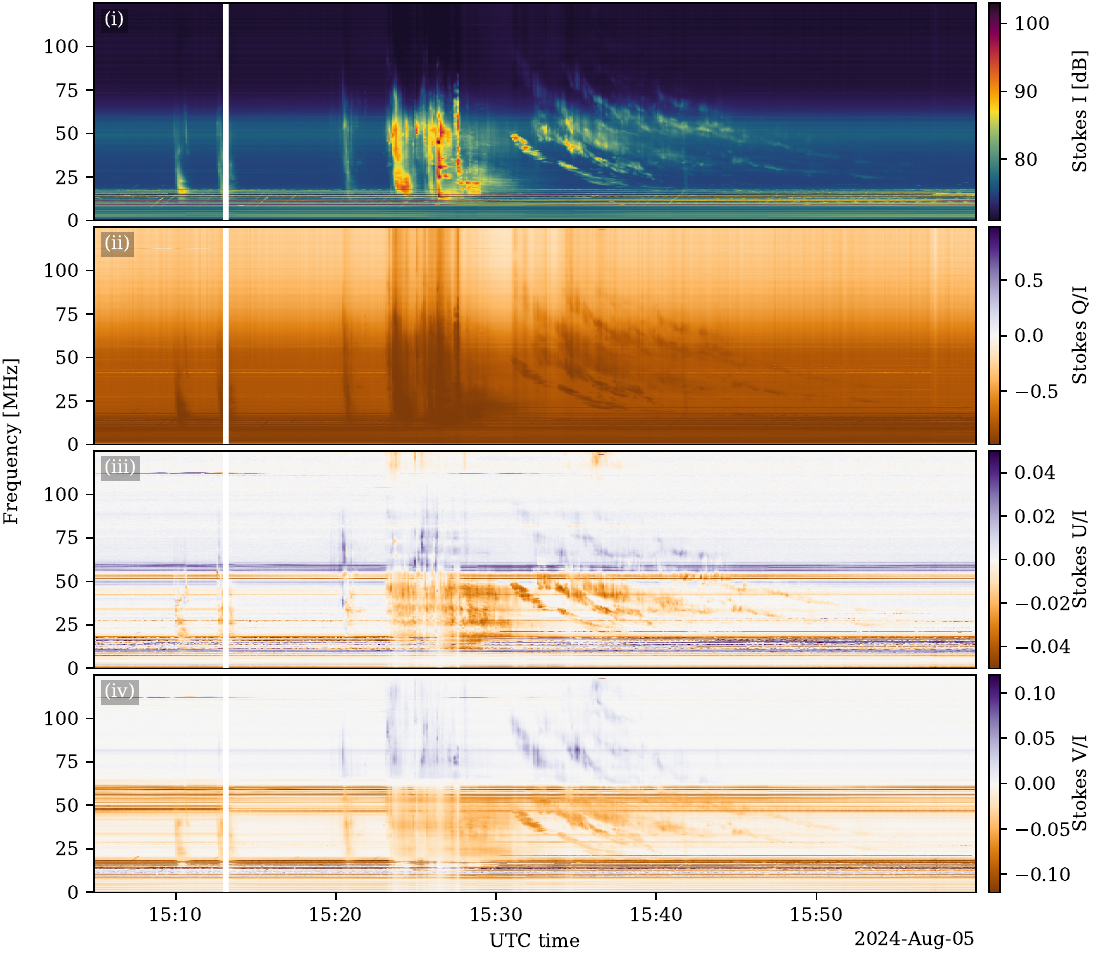}
    \caption{Full-Stokes dynamic spectra of a bright solar radio burst. Panels show Stokes I, Q/I, U/I, and V/I, demonstrating ALBATROS' broadband polarimetric capability. The event exhibits strong frequency-dependent polarised structure in Q/I and weaker but coherent structure in U/I and V/I. Persistent narrowband background below approximately 20~MHz arises primarily from terrestrial high-frequency (HF) radio transmissions. The vertical white stripe is a short data gap.}
    \label{fig:full_stokes}
\end{figure*}

Figure~\ref{fig:full_stokes} presents full-Stokes dynamic spectra of a bright solar radio burst associated with an X1.1-class flare on 2024 August 5. The burst exhibits morphology consistent with a combination of Type III and Type II radio emission, together with rich polarised structure across the observing band. Pronounced frequency-dependent structure is evident in the Stokes Q/I spectrum, while U/I exhibits comparatively weaker coherent features and V/I reaches fractional amplitudes of approximately 10\%. The coherent polarised counterparts to many of the fine spectral structures in Stokes I suggest that the burst morphology is preserved across multiple Stokes parameters. A persistent spectral feature centred near 50--60~MHz is visible throughout the observation and is likely instrumental in origin and will be further characterised through future calibration. 

At frequencies below approximately 20~MHz, the dynamic spectra contain persistent narrowband background from terrestrial high-frequency (HF) radio transmissions. These features are readily distinguished from the transient solar burst, which exhibits broadband spectral evolution, coherent polarised structure, and a close temporal correspondence with the concurrent GOES soft X-ray emission. Together, these observations demonstrate that broadband solar spectropolarimetry is already possible with ALBATROS, even in the presence of persistent terrestrial HF background.

\subsection{Array-Wide Consistency}

\begin{figure*}
    \centering
    \includegraphics[width=\textwidth]{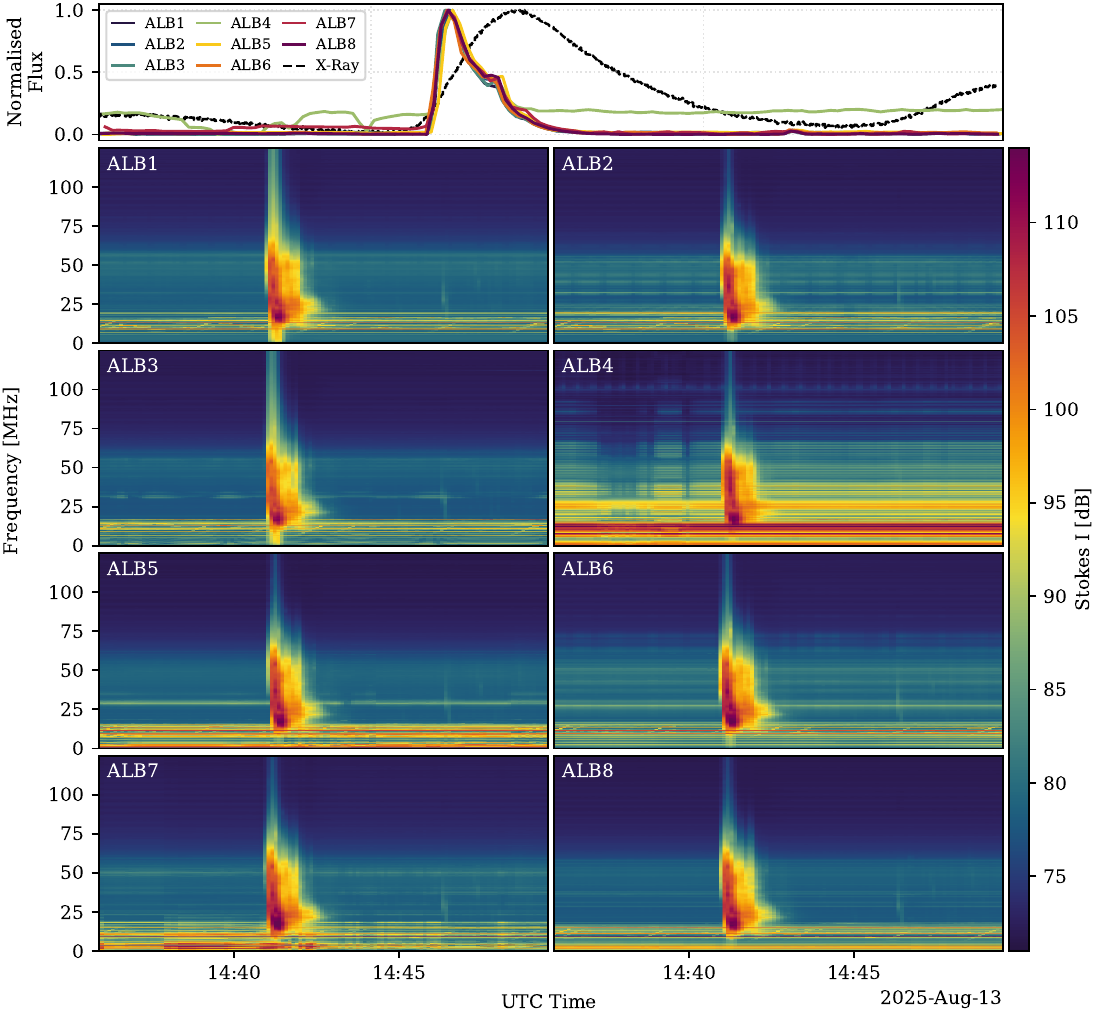}
    \caption{
      Simultaneous observations of the same solar radio burst across all eight ALBATROS stations. The top panel shows normalised frequency-integrated ALBATROS light curves together with concurrent GOES 1--8 \AA\ soft X-ray flux. The lower panels display dynamic spectra for each station using a common color scale. The close agreement among the radio light curves and their correspondence with the soft X-ray emission demonstrate the stability and consistency of the independently operating stations. Variations in background levels reflect differences in the commissioning status of individual stations.
    }
    \label{fig:all_stations}
\end{figure*}

To assess the stability and consistency of the array, Figure~\ref{fig:all_stations} presents simultaneous observations of the same solar radio burst across all eight ALBATROS stations. The burst is detected at every station despite differences in receiver performance and commissioning status. The common spectral morphology and closely matched normalised light curves demonstrate the consistency of the independently operating stations and confirm that the observed emission is a robust solar signal rather than an instrumental artifact.

\section{Discussion and Future Prospects}
\label{sec:discussion}

These first observations demonstrate ALBATROS's capability to detect bright solar radio bursts across the 1--125~MHz band using broadband dynamic spectroscopy, full-Stokes polarimetry, and simultaneous observations from eight autonomous stations. Together, these results establish the ALBATROS as a new facility for ultra-low-frequency solar radio observations. Unlike global networks such as e-Callisto, which achieve continuous coverage through geographically distributed observatories, ALBATROS provides consistent long-duration observations from a single Arctic site while extending routine ground-based access toward the terrestrial ionospheric cutoff.

The Canadian High Arctic provides a unique environment for low-frequency solar observations. Exceptionally low levels of radio-frequency interference, nearly five months of continuous summer daylight, and a polar ionosphere whose plasma frequency spans an unusually wide range of conditions together enable long-duration solar monitoring from a single site. In particular, periods of exceptionally low ionospheric plasma frequency permit observations at frequencies that are only intermittently accessible from most ground-based observatories, while simultaneous measurements of the polar ionosphere offer new opportunities to investigate the response of the high-latitude ionosphere to solar activity.

The capabilities demonstrated here represent only the beginning of ALBATROS’s scientific potential. Archived baseband voltages will enable offline correlation and long-baseline interferometric imaging of solar radio bursts, allowing burst sources to be localized and tracked across the low-frequency corona. The unique combination of long baselines and routine access to frequencies near the terrestrial ionospheric cutoff positions ALBATROS to deliver high-resolution imaging in a spectral regime that remains largely inaccessible to ground-based observatories. The same voltage data also provide access to sub-microsecond time resolution, opening new opportunities to investigate the fine temporal structure of coherent solar radio emission. Following completion of the eight-station array in 2025, ongoing calibration efforts are focused on establishing an absolute flux scale, characterising the sky response, and enabling quantitative polarimetric analyses. These developments will expand ALBATROS’s scientific capabilities, enabling detailed studies of solar radio bursts, the low-frequency corona, and the polar ionosphere.

\section*{Acknowledgements}

We acknowledge the Polar Continental Shelf Program for providing funding and logistical support for the ALBATROS research program. The authors thank Chris Omelon, Laura Thomson, and all the researchers at the McGill Arctic Research Station for their constant support during our field campaigns. We acknowledge the support of the Natural Sciences and Engineering Research Council of Canada (NSERC), RGPIN-2019-04506, RGPNS 534549-19; National Geographic Society Explorer Grant NGS-94983T-22. We acknowledge the support of the Government of Canada’s New Frontiers in Research Fund (NFRF) NFRFE-2023-00069. This research was undertaken, in part, thanks to funding from the Canada 150 Research Chairs Program. A.C. is grateful to Michael Wilensky for helpful discussions and acknowledges support from the Trottier Space Institute Fellowship program. M.A. acknowledges support from Vanier Canada Graduate Fellowship.

\section*{Data Availability}

The data underlying this article will be shared on reasonable request to the corresponding author.


\bibliographystyle{mnras}
\bibliography{solar}


\bsp	
\label{lastpage}
\end{document}